\documentclass[aps,onecolumn,superscriptaddress,12pt]{revtex4}
\usepackage{graphicx}
\usepackage{amsmath}
\begin{document}

\title{Unexpected crossover dynamics of single polymer in a corrugated tube}

\author{Andres De Virgiliis}
\affiliation{Instituto de Investigaciones Fisicoquimicas Teoricas y Aplicadas
(INIFTA), C. C. 16, Suc. 4, 1900 La Plata, Argentina}
\author{Lukasz Kuban}
\affiliation{Institute of Thermal Machinery, Czestochowa University of
Technology, Armii Krajowej 21, 42200 Czestochowa, Poland}
\author{Jaroslaw Paturej}
\affiliation{Institute of Physics, University of Szczecin, Wielkopolska 15,
70451 Szczecin, Poland}
\affiliation{Max-Planck Institut f\"ur Polymerforschung, Ackermannweg 10, 55128
Mainz Germany}
\author{Debashish Mukherji}
\affiliation{Max-Planck Institut f\"ur Polymerforschung, Ackermannweg 10, 55128 Mainz Germany}

\date{\today}


\begin{abstract}
We present molecular dynamics study of a generic (coarse-grained) model for single-polymer
diffusion confined in a corrugated cylinder.
For a narrow tube, i.e., diameter of the cylinder $\delta < 2.3$,
the axial diffusion coefficient $D_{||}$ scales as $D_{||} \propto N^{-3/2}$, with chain length $N$,
up to $N \approx 100$ then crosses over to Rouse scaling for the larger $N$ values.
The $N^{-3/2}$ scaling is due to the large fluctuation of the polymer chain along its fully stretched
equilibrium conformation. The stronger scaling, namely $N^{-3/2}$, is not observed for an atomistically smooth
tube and/or for a cylinder with larger diameter.
\end{abstract}

\maketitle

The dynamics of polymer chains in high geometric confinements has attracted
increased attention over the last decades~\cite{patel89,eisenriegler,site02prl}.
While early studies striven to achieve a fundamental understanding of the
often exotic and unpredictable behavior of polymers in confinement,
recent research has mostly been motivated by applications that range
from biology to tribology~\cite{granick03,muser00,bakajin98prl,reisner05prl}.
Despite significant progress within the field, in particular on
the structural and thermodynamic properties of polymers in confinement,
their dynamics have remained poorly understood.
In particular, the studies are mostly done for strongly adsorbed
polymers~\cite{maier99,sukhishvili00,zhao04jacs,zhang05,
mukherji06,mukherji07mac,mukherji08prl} and/or polymers confined
between two surfaces~\cite{brochard77jcp,milchev98,chen04pre,balducci06macro}.
In this context, both the chain length dependent mobility~\cite{maier99,sukhishvili00}, as well as the effect of varying
surface coverage concentration are studied~\cite{zhao04jacs}.
Furthermore, the studies of these systems are not only restricted to the synthetic polymers~\cite{sukhishvili00,zhao04jacs},
but also to biologically relevant systems, such as DNA, proteins and
phospholipids~\cite{maier99,zhang05,chen04pre,balducci06macro}.

Despite significant interest in studying the polymer dynamics in confinement, a lesser
investigated system is the polymers in cylindrical confinement.
In this context, one of the pioneering works to study the
statics and dynamics of polymer confined in a cylinder employed Monte-Carlo simulations~\cite{kremer84jcp}.
Studies also include the investigation of static properties, such as swollen to globule transition,
in a soft tube~\cite{chen07prl} and the relaxation dynamics of polymer chains confined in a cylinder~\cite{arnold07}.
In a more recent work, it was shown that the DNA confined in a nanotube can exhibit large
fluctuations at the shorter length scales and these fluctuations are ceased at longer
length scales \cite{carpenter11apl}.
Furthermore, as far as the scaling laws of the chain length dependent dynamics is concerned,
axial dynamics in a confining tube is usually said to follow Rouse dynamics, i.e., diffusion coefficient
$D_{||} \propto N^{-1}$ with the degree of polymerization $N$ \cite{milchev94}.
In this context, the computational studies have usually employed ideally flat confining surfaces.
However, what has not (yet) been investigated is the extent to which the corrugation or surface roughness
affects the polymer dynamics in a cylindrical confinement.
Corrugation, more generally speaking breaking translational invariance,
is a necessary ``ingredient'' to exert shear force during the polymer dynamics
in high geometric confinements. In this context, simulating generic bead-spring
polymers moving on surfaces that only had atomic-scale corrugation~\cite{mukherji06,mukherji07mac},
it was shown that the model could reproduce perhaps the counter-intuitive experimental work
of lateral dynamics of polyethylene glycol (PEG) on fused silica surface
by Zhao and Granick~\cite{zhao04jacs}, in which the lateral dynamics $D$ first increased with
the surface coverage concentration $\Gamma$ and then suddenly dropped to a small value at a
threshold concentration $\Gamma^*$. The simulations suggested that this change in $D$ may be due to a
structural transition from single to double layers and that the double layers have more geometric
flexibility to lock into the substrate's registry, which increases energy barriers and thus reduces
lateral mobility at large $\Gamma$~\cite{mukherji06,mukherji07mac}.

In a previous Letter \cite{mukherji08prl}, one of us have shown that the lateral diffusion $D$ of a strongly adsorbed polymer chain
onto a corrugated surface gets strongly influenced by the surface roughness.
The simulations showed $D \propto N^{-3/2}$ with chain length $N$, if the polymer is incommensurate with the substrate
which is consistent with the experiments of polymer adsorbed on a solid surface \cite{sukhishvili00}.
Furthermore, Rouse scaling could be found if the polymer is commensurate with the substrate and/or
for a smooth surface, as observed for polymer dynamics on a soft surface \cite{maier99}.
The weaker scaling is a result of vanishing friction correlation for a polymer adsorbed on a
commensurate or on a smooth surface.

Motivated by the success of our previously used model~\cite{mukherji06,mukherji08prl}, which could reproduce the generic features of the
existing experiments \cite{maier99,sukhishvili00,zhao04jacs}, we revisit the system of polymer dynamics
under cylindrical confinement and study the effect of atomic level corrugation on axial diffusion $D_{||}$.
More specifically, we will investigate the scaling of $D_{||}$ in the presence of surface roughness.

\begin{figure}[ptb]
\includegraphics[width=0.46\textwidth]{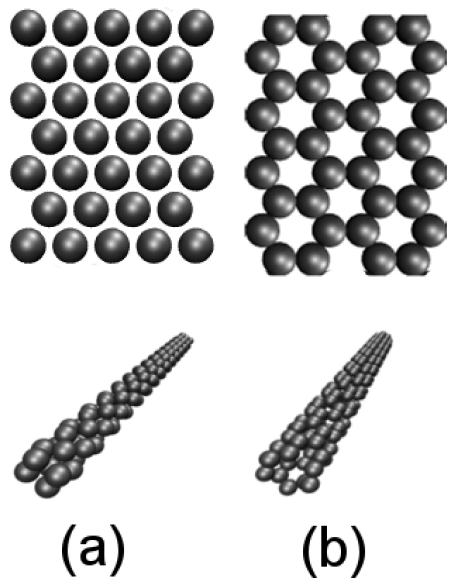}
\caption{(color online) Snapshot of the surface architecture used in the 
simulations, part (a) represents a triangular
lattice and part (b) is a replica carbon nanotube.
\label{fig:snap}}
\end{figure}

In this work, we use the well known bead-spring model ~\cite{kremer90jcp}.
The individual monomer units interact via a truncated repulsive Lennard-Jones (LJ) potential,
with a short range cut-off $r_{\rm c} = 2^{1/6}$.
This $r_{\rm c}$ is chosen to model good solvent condition as well as
potential is continuous at $r_{\rm c} = 2^{1/6}$. Results are presented in terms of
the LJ length scale $\sigma$, the LJ energy parameter $\varepsilon$ and the mass of individual monomer $m$.
Adjacent monomers in the polymer chain interact via an additional Finitely Extensible Nonlinear Elastic (FENE) potential,
\begin{equation}
V_{\rm chain}(r) = \left\{ \begin{array}{cc} - \frac{1}{2} k {R_{0}^2}
\ln [1-(\frac {r}{R_{0}})^2], & \textrm{for $ r {\leq} R_{0}$} \\
\infty & \textrm {elsewhere}
\end{array} \right.
\end{equation}
where $R_0 = 1.5$ and $k = 30$.
The parameters of the potential (i.e., LJ $+$ FENE) ensure no unphysical bond crossing
is allowed and give rise to an effective bond length of around $0.97$.
The bead-spring polymer chain is confined in a tube that has atomic scale roughness.
For this purpose we choose couple of surface architectures. In one case, we choose a surface that has the same
inherent structure as a carbon nanotube and in the second case we roll a triangular lattice into a
cylinder. The surface structures and the corresponding cylinders are presented in the Fig.~\ref{fig:snap}.
Periodic boundary conditions are applied only along the axis of the tube,
i.e., $z$ direction. Monomer and tube atoms interact via a repulsive LJ potential with a
cutoff $2^{1/6}$. The sizes of the polymer chains are varied from $N=20$ to $N=500$.
We choose several different tube diameters $\delta$ in both the systems (we will provide details at the appropriate captions).
For every $N$, the length of the tube is chosen at-least twice the size of the fully stretched polymer conformation.
To avoid commensurability effects \cite{mukherji08prl}, the nearest neighbor atomic distance is chosen to be 0.8
for carbon nanotube architecture and 1.209 for triangular lattice. To mimic static free energy barrier we keep the surface atoms frozen. 
Moreover, it is yet important to mention that we have also conducted a few test runs where the surface atoms were coupled to their respective 
lattice sites with a harmonic spring with a spring constant $\kappa = 5000$. In this case also the results were found to be 
the same as the case of rigid surface atoms.

Temperature is imposed through a Langevin thermostat. Since the substrate atoms are 
constrained to their equilibrium positions, the motion of the polymers needs to be 
thermostated so that constant temperature conditions can be mimicked. 
For this purpose, we chose a Langevin thermostat, which only acts on the motion normal to the axis 
of the cylinder, i.e., normal to the polymer axial dynamics. The coupling strength is chosen as 
$\gamma = 0.001$, which ensures a diffusive behavior within the choices of $\delta$. 
This value of $\gamma$ is chosen after calculating diffusion coefficients for different $\gamma$'s 
and making sure that the diffusion coefficient 
does not get affected by the specific choice of $\gamma$ (data not shown).
Unless stated otherwise the thermal energy is set to $0.5$.
It is yet important to mention that, since our calculations are performed in thermal equilibrium and thus in
linear response, damping due to wall friction and damping
due to hydrodynamic interactions are linearly additive. Therefore, under extreme confinement, as in our case, 
the surface damping will dominate and will screen out the hydrodynamic damping. Hence, we abstain to include explicit 
solvent in order to emphasize more on the wall friction. However, in some cases, influence of hydrodynamic 
interactions can be visible on the lateral dynamics of polymers near smooth surfaces, where hydrodynamic scaling, namely $N^{-3/4}$, 
can be observed \cite{hoda09pre}. Furthermore, as expected, hydrodynamic interactions are screened for a rough surface \cite{hoda09pre}.
In this study, the configurations are equilibrated for a few $10^7$ MD
time steps (depending on the chain size) and then observations
are carried out over another $20 \times 10^6$ MD time
steps; this corresponds to a time of $10^5$ in the LJ units.

\begin{figure*}[ht]
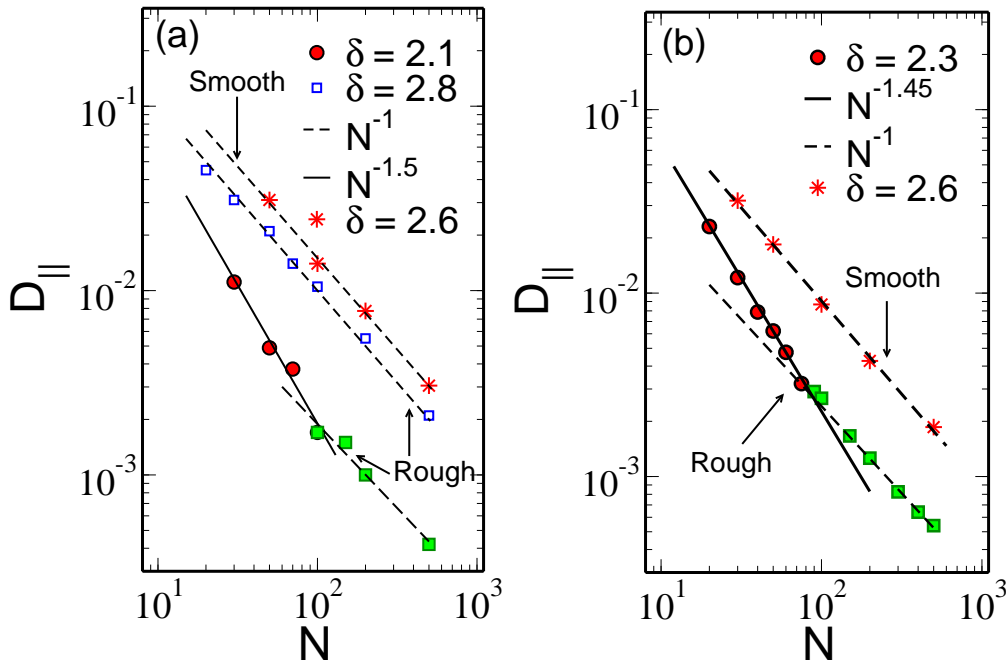

\begin{center}
\includegraphics[width=0.4\textwidth]{diff_N_1stset.eps}
\includegraphics[width=0.4\textwidth]{diff_N_2ndset.eps}
\end{center}
\caption{(color online) Diffusion coefficient $D_{||}$ along the tube axis as a function of chain length $N$. Part (a) shows
the data for the cylinder with carbon nanotube architecture for two different tube diameters $\delta$.
The thermal energy is elevated to $0.7$ for $\delta = 2.1$. We show the results for the default system (rough surface) as well as the result for a smooth surface where the LJ
radius between monomer and surface atom interaction was increased to 1.5. In the case of smooth surface, we choose $\delta = 2.6$ to have the same 
accessible are for the monomer as the case of rough surface and $\delta = 2.1$. In part (b), we show the results for cylinder with
triangular structure. Diameter was chosen as $\delta = 2.3$ and the thermal energy is elevated to $0.7$. Note that the data sets in
two plots are obtained from two different (independent) molecular dynamics softwares. For part (a), we used DL-POLY \cite{dlpoly} and for the part (b) 
we used a self-written parallel MD code.
\label{fig:diff_N}}
\end{figure*}

The main goal of this paper is to study the dynamics of polymers under cylindrical confinement.
However, before describing dynamics we also want to ensure that our model reproduces correct scaling
for the static properties. A quantity that best describe the structural property is the radius of gyration $R_{\rm g}$
(data not shown). We find the radius of gyration along the axis follow $R_{\rm g ||} \propto N$. Furthermore,
for $\delta \le 2.3$ the monomer fluctuations normal to the tube axis are almost negligible,
which ensures an almost one dimensional polymer conformation,
giving a hint for the single file motion in a tube.

Now we focus on the polymer dynamics, in particular on the axial diffusion
coefficients $D_{||}$. For this purpose, we start with the calculation of the
mean square displacement of the individual monomers $C_1(t)$,
\begin{equation}
C_1(t) = \frac{1}{N}
\sum_{i=1}^{N}
\left\langle
\left[
R_{i}(t+t')-R_{i}(t')
\right]^2
\right\rangle,
\label{eq:c_of_t}
\end{equation}
and the center-of-mass $C_2(t)$,
\begin{equation}
C_2(t) =
\left\langle
\left[
\bar R(t+t')-\bar R(t')
\right]^2
\right\rangle,
\label{eq:cm_of_t}
\end{equation}
where $R_{i}$ is the coordinate of the $i^{th}$ monomer along the tube and $\bar R$ is
the center-of-mass coordinate of the polymer chain along the tube.
The lateral diffusion coefficient is related to the $C(t)$, which shows a linear time dependence at
larger times, i.e., $2dD = \lim_{t\to \infty} dC(t)/dt$ with $d$ as the dimensionality.
For monomer mean square displacement, we observe an initial ballistic region where $C_1(t) \propto t^2$
when $t < 1$, then between $1< t < 10^2$ the system exhibits a sub-diffusive behavior
with $C(t) \propto t^{1/2}$. Finally, for $t > 10^2$, the system goes to the diffusive region. This behavior is observed for
$N=100$. However, for larger $N$ values, diffusive region starts at a bit longer time scales.
On the other hand, $C_2(t)$ reaches the diffusive region much earlier
than the monomers and directly crosses over from the ballistic to the diffusive regime.
For all the chain lengths $N$, the $D_{||}$ was calculated by
measuring the slope of $C_2(t)$ at large times.

In the Fig.~\ref{fig:diff_N}, we summarize the most important result for $D_{||}$
for both systems. In part (a) we present the results for replica carbon nanotube and in part (b)
we have the system with triangular geometry. The data for $\delta = 2.8$ in part (a) of Fig.~\ref{fig:diff_N}
are consistent with the scaling law $D \propto N^{-1}$ predicted by Rouse model and with prior simulation
results \cite{milchev94}. It is also important to mention that dynamics of polymer between
two surfaces usually supports the Rouse scaling \cite{mukherji}.
However, upon decreasing the inter-wall distance to a value where the polymer
conformation is ideally flat, one can expect a stronger scaling, namely  $D \propto N^{-3/2}$,
similar to that of the scaling law obtained for the strongly adsorbed polymers \cite{mukherji08prl}.
Though, by no means, we want to make any comparison of cylindrical confinements with the
adsorbed and/or confined polymer cases. The apparent deviation from the Rouse scaling invokes
the question of whether a $N^{-3/2}$ scaling can also be observed in the case of a polymer
confined in a corrugated cylinder.

In order to investigate the effect of corrugation and possible observation of stronger scaling, the
diameter of the confining tube is reduced to a value of $\delta = 2.1$. As indicated earlier,
the polymer conformation is almost one dimensional for $\delta = 2.1$, similar to that of single file
motion. We also want to point out that due to the ``physical" presence of the surface atoms, actual
accessible space for the monomers normal to the surface is around unity in LJ units, which is also
the size of the monomers. Indeed a scaling law of $D \propto N^{-3/2}$ is observed for
$N \le 100$. However, for $N > 100$, interesting enough, we see an ``unexpected" crossover to $D \propto N^{-1}$ (see
Fig.~\ref{fig:diff_N}(a)). Furthermore, in order to check the correctness of this cross-over
dynamics, we run independent simulations on another surface architecture where the surface topography was chosen to be
of triangular lattice. In Fig.~\ref{fig:diff_N}(b), we show the results for $D_{||}$ as a function of $N$. Here we choose
$\delta = 2.3$. Indeed, a cross-over from $D \propto N^{-3/2}$ to Rouse scaling is again observed. Furthermore, we 
see that cross-over occurs at around $N = 100$. (Note: two distinctly different softwares are used to 
obtain these two sets of simulation data).
While the cross-over scalings are always of great interest, it is yet imperative to
understand the molecular origin of the behavior, which is directly accessible in
molecular dynamics simulations.
As of the $N^{-1}$ scaling at large values of $N$, an expected argument would be that
polymers may move as uncorrelated domains and thus lead to Rouse
dynamics. Since a polymer chain is in a single file conformation,
the polymer is thus ‘‘rubbing’’ against the tube where almost every monomer
is in contact with the tube wall, so that the damping (i.e., the inverse diffusion constant)
of the polymers centroid is simply proportional to the number of monomers in contact with
the cylinder. Therefore, for strong confinements the argument naturally leads to Rouse dynamics.

Now we look into the $N^{-3/2}$ scaling: first intuition would suggest
that the motion in this region is correlated and therefore leads to a stronger scaling.
Indeed while looking at the simulation snapshots, we see a
fluctuating motion of the center-of-mass of the polymer around its
stretched equilibrium configuration along the tube axis (similar to that of a trajectory of the
periodic motion in a moving vehicle as observed from a stationary reference frame).
On the other hand the motion is rather smooth for $N > 100$, as expected by the observation of Rouse motion
in this regime.

\begin{figure}[ptb]
\includegraphics[width=0.49\textwidth]{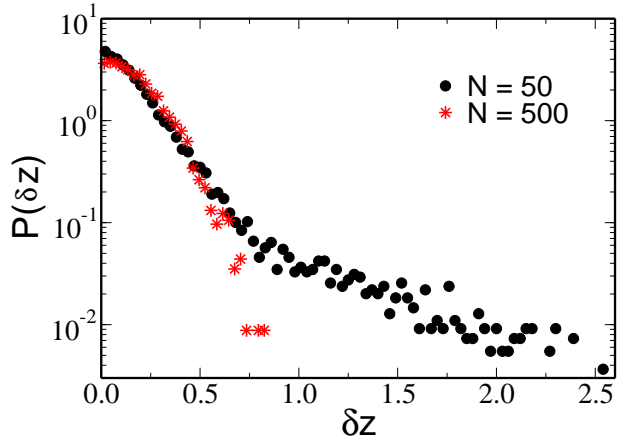}
\caption{(color online) Normalized probability distribution $P(\delta z)$ of the axial hopping distance $\delta z$ of center-of-mass
after every 10 LJ time units for $N = 50$ and 100 LJ time units for $N = 500$, respectively.
\label{fig:cm_t}}
\end{figure}

To make a more quantitative illustration
we have drawn Fig.~\ref{fig:cm_t}, where we show the probability distribution $P(\delta z)$ of the axial hopping distance $\delta z$ of center-of-mass
after every 10 LJ time units for $N = 50$ and 100 LJ time units for $N = 500$, respectively. It can be appreciated that the significant
hopping is observed in the centroid motion for $N=50$, which is visible from the long tail in the probability distribution function.
It is yet important to mention that the fluctuations can be as big as 2 to 3 lattice distances.
Whereas the fluctuations are rather small and/or minimal for the large $N$'s given the large time interval over which the hopping distance is
calculated. These fluctuations enhance the overall correlation along the chain and hence leading
to $N^{-3/2}$ scaling law in the mobility. We have also measured the end-to-end relaxation $\tau$ as a function of chain length. As expected, $\tau \propto N^3$
for Rouse scaling and $\tau \propto N^{3.5}$ for $D \propto N^{-3/2}$.

We also want to emphasize that the fluctuations are observed due to the coupling of transverse
and axial motions in high geometric confinement, which are ``only" due to the presence of atomic scale roughness. This affects the
dissipation of kinetic energy associated with the polymer’s center-of-mass motion along the axis.
Therefore, if our scenario is correct, it would be important to validate if the said $N^{-3/2}$ scaling would disappear
by employing an atomistically flat cylinder and/or under moderate confinement. As of the later case, we have already shown that
$D \propto N^{-1}$ for $\delta = 2.8$ (see Fig.~\ref{fig:diff_N}(a)). 
To mimic the smooth surface we have increased the LJ radius between monomer and surface to 1.5.
We also choose larger $\delta$ that ensures same normal accessible space for the monomer in the smooth tube as that in the case of the rough tube.
As shown in Fig.~\ref{fig:diff_N}(b), data clearly reveal a $N^{-1}$ scaling law. Also, as expected, fluctuations are not observed for the smooth surfaces.

In conclusion, using molecular dynamics simulations of a generic model, we have studied the
dynamics of polymer chain trapped in a ``replica" carbon tube and a corrugated tube with triangular geometry.
We introduced atomic scale roughness to take account of the friction between polymer and cylindrical tube.
We observe a cross-over scaling law for the axial dynamics $D_{||}$ of the polymer chain as a function of chain length $N$,
namely $N^{-3/2}$ to $N^{-1}$. We argue that the cross-over happens due to the surface corrugation that
leads to the coupling of tranverse and axial motions, leading to large axial fluctuations of the center-of-mass for
chain lengths $N< 100$. Whereas, the fluctuations are ceased for the longer chains and hence the Rouse scaling is observed.
The cross-over disappears for the tube diameter longer than the monomer size and/or for a smooth surface.
The results presented in this work might be of particular importance for the study of the translocation of 
polymers through a narrow channel. In this case, the ``so called" scaling behavior of the translocation time $\tau$ will
strongly depend on the dynamics within the cylindrical confinement \cite{paturej}.

We thank Dominik Fritz and Jia-Wei Shen for critical reading of the manuscript. 
Computational time on PL-Grid Infrastructure is greatfuly acknowledged.

\end{document}